\documentclass[aps,pre,twocolumn,superscriptaddress]{revtex4}

\usepackage{graphics}

\begin{document}


\title{Congestion-gradient driven transport on complex networks}


\author{Bogdan Danila}
\email[]{dbogdan@mail.uh.edu}
\affiliation{Department of Physics, The University of Houston, Houston TX 77004}

\author{Yong Yu}
\affiliation{Department of Physics, The University of Houston, Houston TX 77004}

\author{Samuel Earl}
\affiliation{SI International, Beeches Professional Campus, Rome NY 13440}

\author{John A. Marsh}
\affiliation{SI International, Beeches Professional Campus, Rome NY 13440}

\author{Zolt\'an Toroczkai}
\altaffiliation{At Department of Physics, University of Notre Dame,
Notre Dame IN 46556, from June 1st, 2006}
\affiliation{Los Alamos National Laboratory, MS B258, Los Alamos NM 87545}

\author{Kevin E. Bassler}
\email[]{bassler@uh.edu}
\affiliation{Department of Physics, The University of Houston, Houston TX 77004}

\date{\today}

\begin{abstract}
We present a study of transport on complex networks with routing based on local information. Particles hop from one node of the network to another according to a set of routing rules with different degrees of congestion awareness, ranging from random diffusion to rigid congestion-gradient driven flow. Each node can be either source or destination for particles and all nodes have the same routing capacity, which are features of ad-hoc wireless networks. It is shown that the transport capacity increases when a small amount of congestion awareness is present in the routing rules, and that it then decreases as the routing rules become too rigid when the flow becomes strictly congestion-gradient driven. Therefore, an optimum value of the congestion awareness exists in the routing rules. It is also shown that, in the limit of a large number of nodes, networks using routing based on local information jam at any nonzero load. Finally, we study the correlation between congestion at node level and a betweenness centrality measure.
\end{abstract}

\maketitle

\section{Introduction}
\noindent Network transport has been a topic of intensive research in recent years \cite{NewmanSIAM,WattsStro,HolmeCong,EcheniquePRE,EcheniqueEL,Anghel-ea,Lopez-ea,Korniss-ea,Guimera} due to the wide variety of network systems of practical interest, both natural and human-made. These include \cite{NewmanSIAM,BarabRMP,BarabLinked} the Internet, the world wide web, wireless communication networks, networks of biological processes, social networks, and various distribution networks that make up the infrastructure of technological society. Research topics include the robustness and efficiency of computer and distribution networks, the spread of disease, and the study of the interplay between various biological or economic processes.

A major result obtained recently \cite{NewmanSIAM,BarabRMP,BarabLinked,NewmanPRE} is the realization that both natural and man-made networks often display scale-free topology, i.e.\ they are characterized by a power-law distribution of the node degrees. The explanation for this observation, namely the connection between form and functionality in networks, remains open. There exist a good number of stochastic models for network growth and evolution that lead to scale-free graphs, such as the preferential attachment model of Barab\'asi and Albert. \cite{BarabScience} However, they are not directly related to the transport function of the network. In this paper we explore the possibility of a more direct connection between topology and transport function in complex networks.

Specifically, we explore the dependence of transport capacity on the structure of the network, as well as on the routing rules. We consider a simple transport model, motivated in part on the conditions applicable to a wireless ad-hoc network, \cite{ZJHaas} where any node may be either the source or the destination of information, and in which all nodes have equal routing capacity. We consider a dynamics in which ``particles" (representing information, energy or goods) are routed sequentially from node to node along undirected network edges until they reach their randomly preassigned destination. The routing choices are made based on local information only. Particles are added to the network at a constant rate, with equal probability for every node, and removed upon reaching their destination. The network load is quantified by the particle generation rate. Each individual network is characterized by a critical value of the load \cite{Guimera,ZhaoLaiParkYe} beyond which it enters a jammed state, i.e., the average time to destination diverges and consequently the number of particles on the network increases indefinitely. For networks characterized by a given set of structure and routing parameters, we define the average jamming fraction as the ratio of the number of network realizations that end up jamming divided by the total number of realizations. The critical value of the load at which the average jamming fraction reaches 50\% is used to quantify the network transport capacity from a statistical point of view.

This study is also motivated in part by the work presented in Refs.\ \cite{TB} and \cite{TKBHK}, where the notion of gradient flow networks is introduced. Gradient networks are directed subnetworks of an undirected ``substrate" network in which each node has an associated scalar potential and one out-link that points to the node with the smallest (or largest) potential in its neighborhood, defined as the reunion of itself and its nearest neighbors on the substrate network. Here we use a congestion aware routing rule characterized by a parameter which defines a continuous change from random diffusion to higher degrees of congestion awareness. Increasing congestion awareness results in improved network load balancing. The limiting case of congestion aware routing is that of rigid congestion-gradient driven flow, when particles are always routed towards the least congested neighbor. In this case, transport takes place along a gradient network, but one that changes dynamically in a manner that is correlated with the degrees of congestion of the nodes.

We present results for two network topologies, random and scale-free. Random (also known as Erd\H{o}s-R\'enyi) networks are characterized \cite{Erdos} by the number of nodes and by a constant probability $p$ for the connection between any given pair of nodes. The scale-free networks we study are grown using the Barab\'asi-Albert algorithm of growth by preferential attachment. \cite{BarabScience} When results for the two types of networks are compared in this paper, the number of nodes and the average degree are the same. Recently, a detailed analysis of jamming in gradient networks \cite{ParkLaiZhaoYe} suggested the existence of a critical value of the average degree. For values of the average degree below this critical value, large scale-free networks are somewhat more prone to jamming than random networks with the same number of nodes and average degree while the opposite is true above the critical value.

Our main result is the existence of an optimum value of the congestion awareness parameter, both for random and for scale-free networks. Below this optimum value, the transport capacity of the network increases with the degree of congestion awareness due to the particles being more likely to avoid waiting in the queues of the busiest nodes. However, above the optimum value the average transport capacity decreases with increasing congestion awareness in spite of the shorter waiting times. The decrease is mainly due to the formation of transport traps, which prevent particle flow between parts of the network. Our results also show that, for a value of the average connectivity below the critical value suggested in \cite{ParkLaiZhaoYe}, scale-free networks are indeed more prone to jamming than random networks regardless of the degree of congestion awareness.

The outline of the paper is as follows. In Section II we give a detailed description of our model and of the criteria that were used to detect jamming. In Section III we present results for the jamming fraction as a function of load for different values of the number of nodes and of the congestion awareness parameter. Also in Section III we present results for the critical load as a function of both the number of nodes and the congestion awareness parameter. In Section IV we present an in-depth analysis of the transport behavior. Section V summarizes our results and conclusions.

\section{Model}

\noindent Our model network consists of a set of $N$ identical nodes linked together in such a way that they form a connected, undirected graph characterized by a symmetric adjacency matrix $A$, defined by
\begin{equation}
	A_{ij}=\left\{
	\begin{array}{ll}
	1 & \mbox{if $i$ and $j$ are connected}, \\
	0 & \mbox{otherwise}.
	\end{array}
	\right.
\end{equation}
\noindent The degree $d$ of a node is defined as the number of links (edges) that connect it to other nodes. We have studied both the case of Erd\H{o}s-R\'enyi \cite{Erdos} (random) graphs, characterized by a binomial distribution of the node degrees, and the case of Barab\'asi-Albert \cite{BarabScience} (scale free) graphs, for which the node degrees obey a power-law distribution. In the case of a random network, the average degree is given \cite{Erdos} by $<d>=(N-1)p$, where $p$ is the probability for a link to exist between any given pair of nodes. We discard random networks that turn out to be disconnected. Scale-free networks grown according to the Barab\'asi-Albert algorithm have an average degree given by
\begin{equation}
	<d>(N)=2m\frac{N-N_0+\frac{L_0}{m}}{N},
\end{equation}
\noindent where $N_0$ is the initial number of nodes, $L_0=N_0(N_0-1)/2$ is the initial number of links, and $m$ is the number of links created with every new node. These networks are always connected. The condition for the average degree to be independent of $N$ is $N_0=2m+1$. All results for scale-free networks presented in this paper have been obtained for values of $m=3$ and $N_0=7$, and consequently $<d>=6$. To facilitate comparison, the probability $p$ that characterizes the random networks has been in most cases adjusted to $p=6/(N-1)$. Varying the average degree does not lead to any qualitative changes in the results. However, the transport capacity shows an overall increase with increasing average degree, while its maximum at the optimum value of the congestion awareness parameter becomes less prominent.

Information packets (or any other entities) transported along the network are represented by particles that hop from one node to the next along the graph edges according to rules that will be discussed in the following paragraph. Each node has a particle queue, thus being capable of holding more than one particle at a time. Particles waiting in a queue are processed according to the ``first-in/first-out" rule. Updating of the network is done sequentially, so that only one particle is moving or being created at any given instant. Each time step consists of a random sequence of updating events, in the course of which $N$ nodes are randomly selected one at a time to forward a particle, and a random number of new particles are added to the network one at a time, with an average rate of $R$ new particles per time step. All nodes are equally likely to receive a new particle. Every new particle generated is assigned a destination, again with equal probability from among all nodes, and is placed at the end of the queue of its origin node. Upon reaching their destination, particles are removed from the network.

The particle hopping dynamics is characterized by the routing rule used to choose the next location of a particle sitting on a given node from among its neighbors. If no record is kept of the possible paths towards destination, the routing rule will necessarily be based on local information only, such as the degrees of the neighboring nodes or any other parameters that characterize their congestion status. If, on the other hand, knowledge of some or all possible routes to destination is used in making the decision, the rule is said to be based on global information. Among the routing rules based on local information we distinguish between random diffusion, in which case the next location is chosen with uniform probability distribution from among the neighbors of the current node, and congestion aware rules that take into account the degrees of congestion of the neighbors. The extreme case of congestion aware routing is that of rigid congestion-gradient driven flow, when particles are transported only in the direction of the gradient network generated by using a congestion parameter as scalar potential. \cite{TB} We have chosen a congestion aware routing rule that uses local information, namely the queue lengths of the neighboring nodes. According to this rule, for a particle currently at node $i$ the probability of hopping to node $j$ is given by
\begin{equation}
	P_{ji}=\frac{A_{ji} (q_j+1)^{-\beta}}{\sum_{k=1}^{N} A_{ki} (q_k+1)^{-\beta}},
\end{equation}
\noindent where $q_k$ is the queue length of node $k$ and $\beta$ is an adjustable parameter. Varying $\beta$ from 0 to $\infty$ allows a smooth transition between the case of random diffusion and that of transport along the gradient network generated by using the queue length as scalar potential. In practice, a value of $\beta\approx 10$ is sufficiently high to lead to essentially rigid congestion aware routing.

For a given realization of the network, simulations were run for a maximum of $10^5$ time steps. The simulations were interrupted and jamming declared at any time if either the length of any queue exceeded the maximum value of $20 N$, or the total number of particles $n$ on the network exceeded $50 N$. In addition, a linear regression for $n(t)$ was attempted every $2\times 10^4$ time steps. Jamming was declared if the result of the regression was a positive slope with a Pearson correlation coefficient $r^2>0.9$ over the last interval of $2\times 10^4$ time steps.

\section{Results for the transport capacity}

\noindent In this section we present results for the average jamming fraction $f_j$ computed as a function of the load $R$ as well as for the average transport capacity, quantified by the critical load $R^*$ at which the jamming fraction reaches 50\%. The results for the jamming fraction were obtained by averaging over 1000 realizations of the network for a given set of parameters $R$, $\beta$, and $N$. The critical load is obtained by a linear fit between two points on the $f_j(R)$ curve, one below $f_j=0.5$ and another one above.
\begin{figure*}
	\scalebox{0.6}{\includegraphics{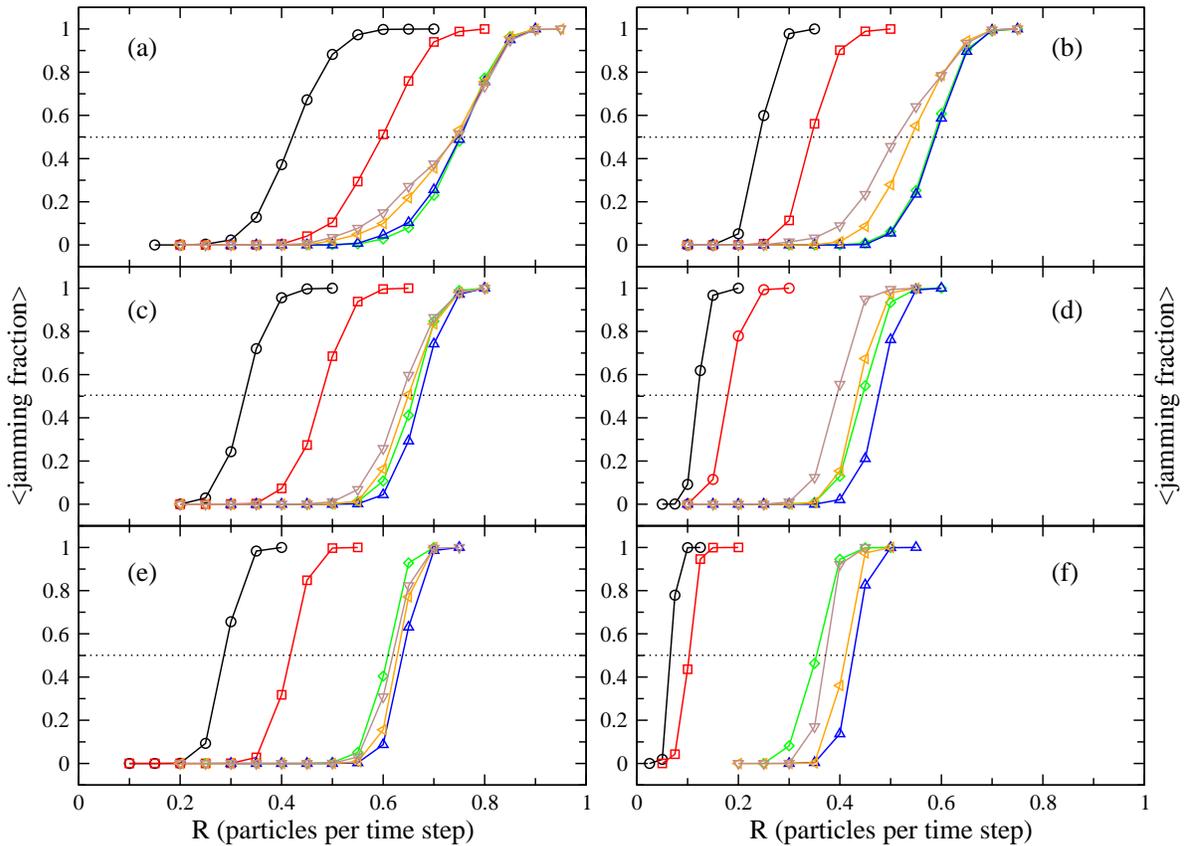}}
	\caption{(Color online) Average jamming fraction as a function of load for $\beta$=0 (circles), 0.1 (squares), 0.5 (diamonds), 1 (up triangles), 3 (left triangles), and 10 (down triangles). Results are for random (a,c,e) and scale-free (b,d,f) networks with $N$=30 (a,b), 100 (c,d), and 300 (e,f) nodes.}
\end{figure*}
\begin{figure*}
	\scalebox{0.45}{\includegraphics{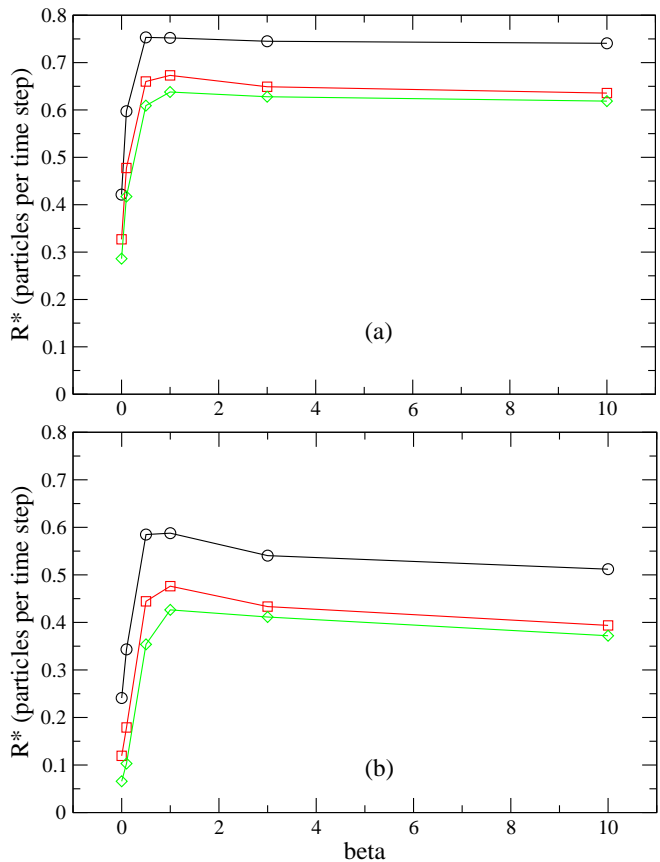}}
	\caption{(Color online) Critical load of the network as a function of the congestion awareness parameter for $N$=30 (circles), 100 (squares), and 300 (diamonds). Results are for random (a) and scale-free (b) networks with $<d>=6$.}
\end{figure*}

Fig.\ 1 shows plots of the average jamming fraction as a function of $R$ for both random (Figs.\ 1a,c,e) and scale-free (Figs.\ 1b,d,f) networks and for values of $N$=30, 100 and 300. Each plot has six curves, corresponding to values of $\beta$=0, 0.1, 0.5, 1, 3, and 10. Jamming of the random networks is seen to occur at higher values of the load than in the case of their scale-free counterparts. This is true regardless of the number of nodes and for all routing regimes, which range from random diffusion ($\beta=0$) to essentially deterministic congestion aware routing ($\beta=10$). The importance of even a slight degree of congestion awareness is also apparent. In most plots, the critical load corresponding to $\beta=0.1$ is almost half-way between the value corresponding to $\beta=0$ and the highest value. Another important conclusion is that, regardless of network topology or routing regime, the critical load decreases as the number of nodes increases. This result is even more significant in view of the fact that the most meaningful quantity is arguably $R/N$, the average number of new particles per node and time step. In the limit of large number of nodes this quantity approaches zero. The significance of this result is that large networks with routing based on local information only will jam for any nonzero value of the load.

Perhaps the most important result is the existence of an optimum value of the congestion awareness parameter $\beta$. This can be seen more clearly in Figs.\ 2 a and b, where the critical load $R^*$ is plotted against $\beta$ for random and scale-free networks respectively. Regardless of network topology or number of nodes, the critical load increases as the congestion awareness parameter increases until it reaches a maximum, corresponding to a value of $\beta$ in the range 0.5 to 1. For higher values of $\beta$ the critical load decreases monotonically and seems to approach a finite value as $\beta\rightarrow\infty$. This seemingly counterintuitive result is explained in Section IV. To facilitate comparison between random and scale-free networks, we present here results for $<d>=6$. However, the maximum of the critical load is more pronounced at lower average connectivities. As the number of nodes on the network increases, the maximum becomes less prominent.

\section{Detailed analysis}

\subsection{Betweenness definition and calculation}

\noindent We begin this section by defining a measure that characterizes the nodes of the network from a purely topological point of view. As we will see, this quantity is a useful tool for understanding the behavior of the transport along the network. Various definitions of betweenness centrality measures have been proposed, each having its scope of applications. \cite{HolmeCong,Guimera,Newman-e,Estrada,Latora-e,Brajendra} The measure that we are using in this paper is a slightly modified version of the betweenness defined by Guimer\`a et al. \cite{Guimera} Random walks along the network are considered for particles with a given source node $s$ and destination node $t$. The betweenness $b_i^{st}$ of a node $i$ with respect to $s$ and $t$ is defined as the average number of times a random walk between $s$ and $t$ passes through $i$. Finally, the global betweenness of node $i$ is defined as the average over all $s$ and $t$ of $b_i^{st}$.

In principle, a calculation of the betweenness can be done by starting from the transition probability matrix $P$, whose elements $P_{ij}$ give the probability for a particle to arrive at node $i$ coming from node $j$. However, the computation of $P$ is straightforward only in the case of random diffusion. Nevertheless, the results presented in the remainder of this section show that, even in the case of congestion aware routing, there exists a strong correlation between the random walk betweenness of a node computed in the case of random diffusion and the average number of particles it receives per time step. Assuming at first that particles are never removed from the network, we have
\begin{equation}
	P_{ij}=\frac{A_{ij}}{d_j},
\end{equation}
\noindent where $d_j$ is the degree of node $j$. For particles with destination node $t$, we can account for their removal once they reach their destination by replacing all elements in column $t$ by 0. The matrix thus obtained will be denoted by $P^{t}$. It is easy to verify that $\left( P^{t}\right)^n_{ij}$, where $\left( P^{t}\right)^n$ is the $n$-th power of $P^{t}$, gives the probability of arrival at node $i$ coming from node $j$ in $n$ steps for a particle with destination $t$. Then the betweenness of node $i$ with respect to a source node $s$ and a destination node $t$ is defined as
\begin{equation}
	b_i^{st}=\sum_{n=0}^{\infty} \left( P^{t}\right)^n_{is},
\end{equation}
\noindent where $\left( P^{t}\right)^0$ is taken to be the unit matrix $I$. Eq. (5) can be rewritten as
\begin{equation}
	b_i^{st}=\left( I-P^{t}\right)^{-1}_{is}.
\end{equation}
\noindent Finally, the global betweenness of node $i$ is given by
\begin{equation}
	b_i=\frac{1}{N^2} \sum_{s,t=1}^N \left( I-P^{t}\right)^{-1}_{is}.
\end{equation}
\noindent If this quantity is multiplied by the rate at which new particles are added to the network $R$, we obtain the average total number of particles (new or from other nodes) reaching node $i$ in the course of a time step.

In its most straightforward form, the algorithm requires $N$ matrix inversions, the number of flops thus being $\mathcal{O}(N^4)$. However, since the transformation from $(I-P^{t})$ to $(I-P^{t'})$ is a rank 2 perturbation of $(I-P^{t})$, one can use the Sherman-Morrison-Woodbury formula \cite{SMWformula} to compute $N-1$ of the inverses at a much lower cost once a first inverse has been computed. Specifically, the computation can be done as follows. First, compute $Q=(I-P^{t})^{-1}$ for a given $t$, for example $t=1$. Then for every $t'\ne t$ we can write
\begin{equation}
	P^{t'}=P^t+\left[P_{:t}|-P_{:t'}\right]\left[\frac{e_t^*}{e_{t'}^*}\right],
\end{equation}
\noindent where $P_{:s}$ denotes the $s$ column of matrix $P$ and $e_s^*$ the $s$ row of the unit matrix. The vertical and horizontal bars within the two bracketed expressions in (8) are column and row delimiters respectively. By an application of the Sherman-Morrison-Woodbury formula to $(I-P^{t'})^{-1}$ using (8), and after a little bit of algebra we find
\begin{widetext}
\begin{equation}
	(I-P^{t'})^{-1}=Q+\left[QP_{:t}|e_{t'}-Q_{:t'}\right]
	\left[
	\begin{array}{ll}
	1-(QP)_{tt} & Q_{tt'} \\
	-(QP)_{t't} & Q_{t't'}
	\end{array}
	\right]^{-1}
	\left[\frac{Q_{t:}}{Q_{t':}}\right].
\end{equation}
\end{widetext}
\noindent Here $e_{t'}$ is the $t'$ column of the unit matrix, while $Q_{:s}$ and $Q_{s:}$ are the $s$ column and respectively row of matrix $Q$. Note that, once $Q$ and the product $QP_{:t}$ have been computed, all coefficients in (9) are known for every $t'$. By an inspection of (9) we find that the total number of flops necessary for the computation of the $N$ inverses becomes $\mathcal{O}(5N^3)$.

\subsection{Correlations with congestion parameters}

\noindent Next we investigate the correlation of the betweenness measure $b$ with two parameters that can be used to characterize the congestion status of a node. Specifically, we look at the correlation with the time averages of the queue length $q$ and particle flux $w$, which is defined as the number of particles received by the node in a time step. The results presented in this subsection pertain only to cases of steady state transport, when the network does not jam. Jamming can be characterized in terms of the particle flux as a situation in which its average exceeds the maximum value of $<w>=1$ for at least one node (all nodes can process on average at most one particle per time step). In addition, since the results are similar regardless of the number of nodes, we only show results for random and scale-free networks with $N=30$ nodes.
\begin{figure*}
	\scalebox{0.6}{\includegraphics{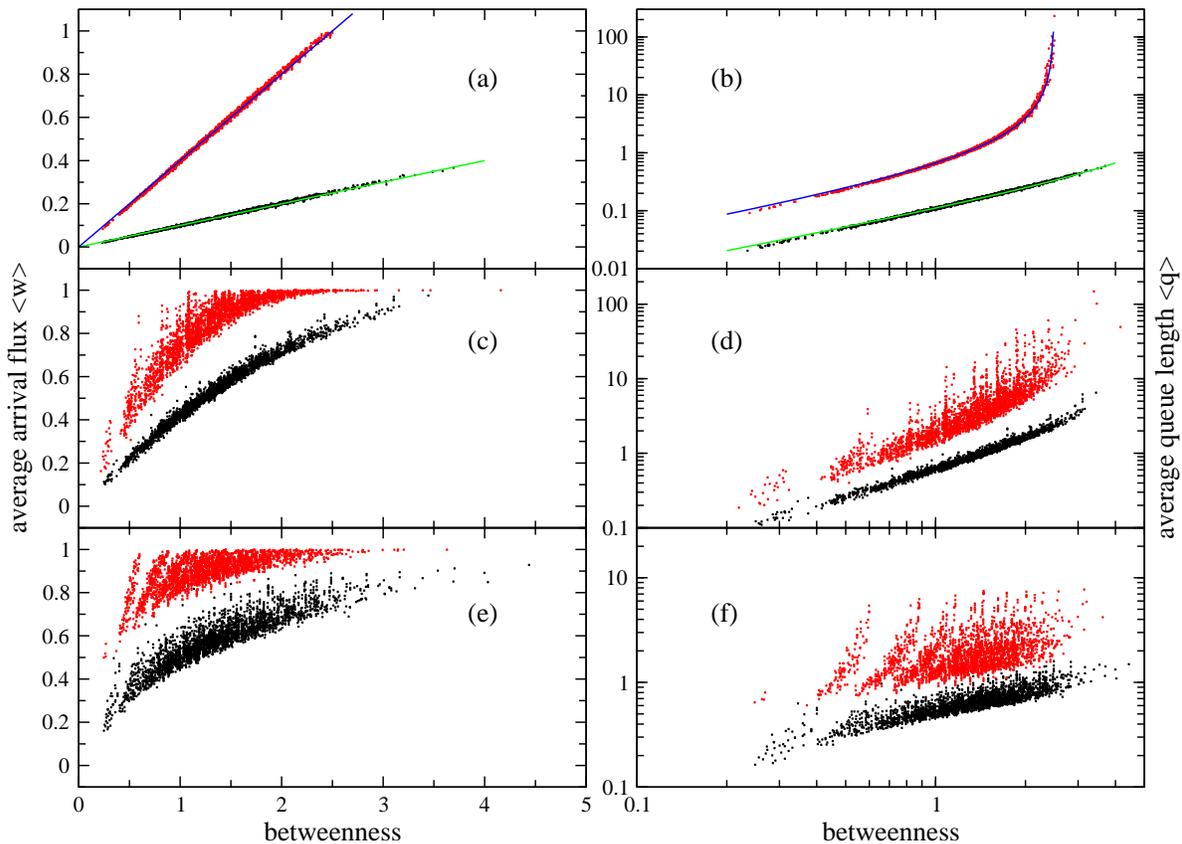}}
	\caption{(Color online) Correlation plots of the average particle flux (left) and average queue length (right) versus betweenness for random networks. The values of $\beta$ are 0 (a,b), 0.5 (c,d), and 10 (e,f). Lower (black) sets of dots are for $R$=0.1 (a,b) and $R$=0.4 (c,d,e,f) while upper (red) sets of dots are for $R$=0.4 (a,b) and $R$=0.65 (c,d,e,f).}
\end{figure*}
\begin{figure*}
	\scalebox{0.6}{\includegraphics{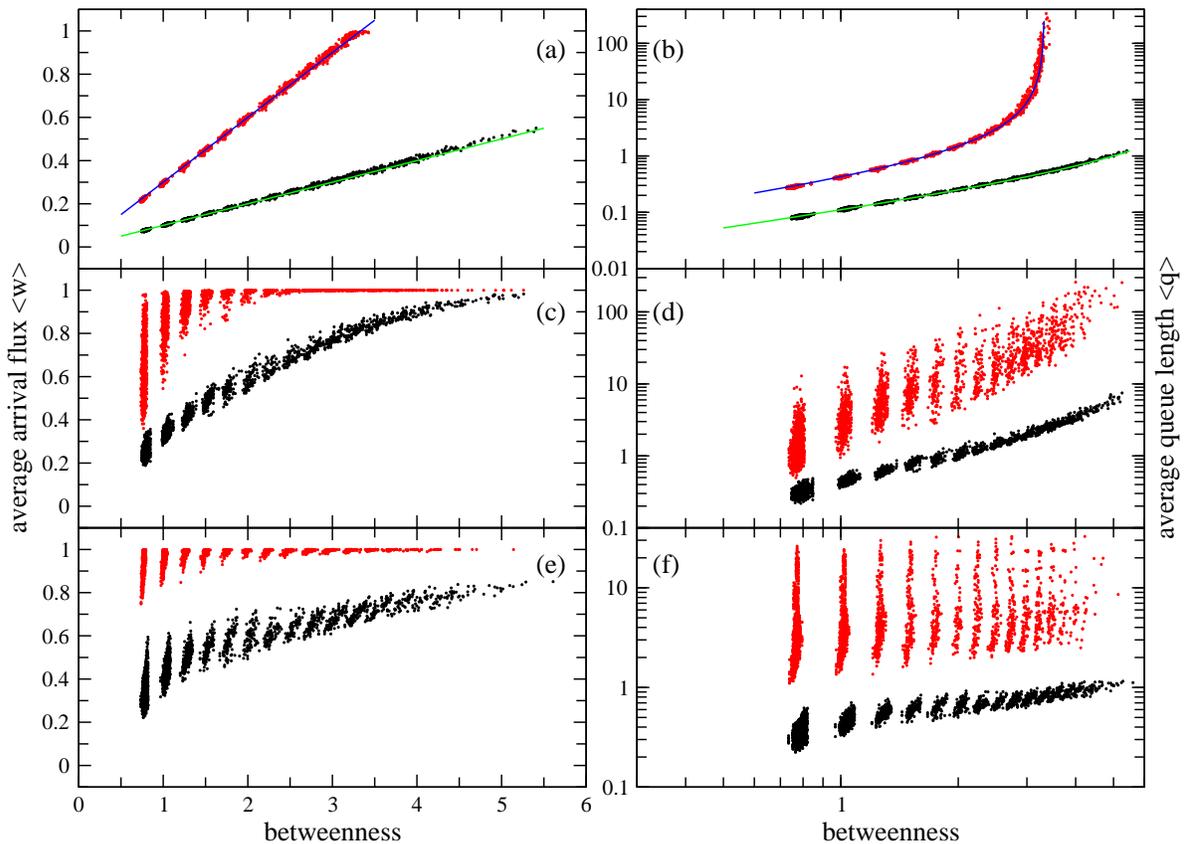}}
	\caption{(Color online) Correlation plots of the average particle flux (left) and average queue length (right) versus betweenness for scale-free networks. The values of $\beta$ are 0 (a,b), 0.5 (c,d), and 10 (e,f). Lower (black) sets of dots are for $R$=0.1 (a,b) and $R$=0.3 (c,d,e,f) while upper (red) sets of dots are for $R$=0.3 (a,b) and $R$=0.65 (c,d,e,f).}
\end{figure*}

Figs.\ 3 and 4 show correlation plots of the average particle flux and average queue length against node betweenness in the case of random and scale-free networks respectively, for values of $\beta$=0, 0.5, and 10. The average particle flux is plotted in Figs.\ 3a,c,e and 4a,c,e while Figs.\ 3b,d,f and 4b,d,f show the average queue length. The averages are computed over $10^5$ time steps. Each plot contains points corresponding to a value of $R$ that is well below $R^*$ (lower set of dots, colored black online) and to a value close to $R^*$ (upper set of dots, colored red online). Each set contains 3000 dots, corresponding to 100 network realizations with a given set of parameters.

As one would expect based on the way the betweenness measure is defined, it provides an essentially exact description of the statistics of the particle transport in the case of random diffusion ($\beta=0$). There is a very strong linear correlation between the average particle flux through a node and its betweenness, regardless of the value of the load $R$. If the points corresponding to a given load are fitted with a straight line the value of the slope is, to within the statistical uncertainty, equal to $R$ (straight lines in Figs.\ 3a and 4a). We thus see that the probability for a network to jam at a given load under random diffusion routing regime is the same as the probability of having at least one node with betweenness higher than the inverse of the load. This also explains the uneven jamming that characterizes networks when the random diffusion routing rule is used, since only the nodes for which $bR>1$ are jamming. As for the average queue length, its correlation with the betweenness measure is approximately linear only when the network load is low, well below its critical value $R^*$. At higher loads, the correlation is still strong but nonlinear. This is due to the fact that the functional dependence of the average queue length on the average particle flux is nonlinear. One can show that, in the case of random diffusion, the number of particles in a queue is distributed according to an exponential distribution law \cite{Guimera,Allen} given by
\begin{equation}
	p(q)=(1-<w>) <w>^q,
\end{equation}
\noindent where $p(q)$ is the probability of having $q$ particles in the queue of a node characterized by an average particle flux $<w>$. From (10) it follows that the average queue length is related to the average particle flux by \cite{Guimera,Allen}
\begin{equation}
	<q>=\frac{<w>}{1-<w>}.
\end{equation}
\noindent The latter formula, with $<w>=bR$, has been used to generate the curves in Figs.\ 3b and 4b. As can be seen, these curves provide an excellent fit for the data points obtained from simulations.

Things are more complicated in the case of congestion aware routing. Neither the average particle flux nor the average queue length correlate linearly with the random walk betweenness except at very low loads. The correlations also become weaker as the load increases. The strength of the correlation is nevertheless surprising, especially in the case of the average flux, and the plots provide some useful insight into the behavior of the network transport. The most meaningful comparison is between the $b-<w>$ correlation plots (Figs.\ 3a,c,e and 4a,c,e). By comparing these figures we see that the way a congestion aware routing rule is able to cope with higher loads than random diffusion is by diverting part of the traffic towards lower betweenness nodes. High betweenness nodes that would otherwise have exceeded the maximum particle flux now are kept just below $<w>=1$, while lower betweenness nodes experience higher traffic. The rigid straight-line correlation between node betweenness and average particle flux is bent, which also explains why networks with a congestion aware routing rule jam more uniformly. The load balancing becomes more pronounced as the load increases. At a given load, the balancing is stronger for higher values of the congestion awareness parameter. One could attempt an explanation for the decrease in the average transport capacity of the network at large $\beta$ in terms of the routing rule ``overshooting" its goal of balancing the network load. Since large values of $\beta$ lead to more rigorous load balancing, the particle flux through the high betweenness nodes is lower than in the case of optimum $\beta$. But high betweenness nodes tend to have high degrees as well and are in general better positioned to route a particle between two weakly connected parts of the network. This could lead to situations where some particles wander about for a longer time before finding their target and, in some cases, to jamming due to the particle flux through some node exceeding its maximum value. To verify this, we looked at the statistics of the target finding times.
\begin{figure*}
	\scalebox{0.4}{\includegraphics{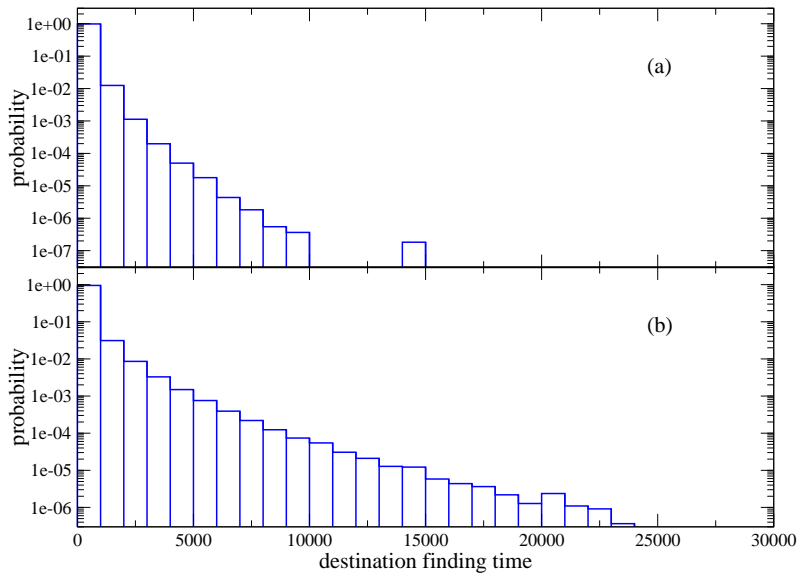}}
	\caption{Histograms of the target finding times at (a) $\beta=1$ and (b) $\beta=10$, for scale-free networks with $N=30$ nodes at $R=0.55$.}
\end{figure*}

In Fig.\ 5 we present histograms of the target finding times at $\beta=1$ and $\beta=10$ for scale-free networks with $N=30$ nodes. The network load is $R=0.55$ (close to $R^*$ for both values of $\beta$) and the statistics is based on 100 non-jamming realizations. It is apparent from Fig.\ 5 that, even in the case of steady state transport, there are particles with much longer finding times at $\beta=10$ than at $\beta=1$. This is an indication that a routing rule too rigidly based on congestion awareness may fail to achieve its goal of minimizing the finding times in the case of jamming networks. The average time a particle sits in a queue is shorter, but this may be compensated by the lengthening experienced by the particle routes. However, the generalization of this explanation to situations of network jamming is not straightforward and, as we will see in the next subsection, it does not provide a complete picture of the jamming mechanisms.

\subsection{Jamming scenarios}

\noindent A more detailed analysis of the transport parameters for jamming networks reveals that the main cause of the decrease in transport capacity at large values of $\beta$ is the onset of transport traps that involve at least one very low degree (and thus low betweenness) node. These traps effectively break the network into two or more subnetworks that are disjoint from the point of view of particle transport. Consequently, particles generated on one subnetwork whose destination happens to be on another one will never be able to find their destination. There are also cases when, due to statistical fluctuations, the subnetworks are not disjoint all the time. However, jamming will occur if the time intervals when they can exchange particles are on average too short to allow all particles to find their destination.
\begin{figure*}
	\scalebox{0.3}{\includegraphics{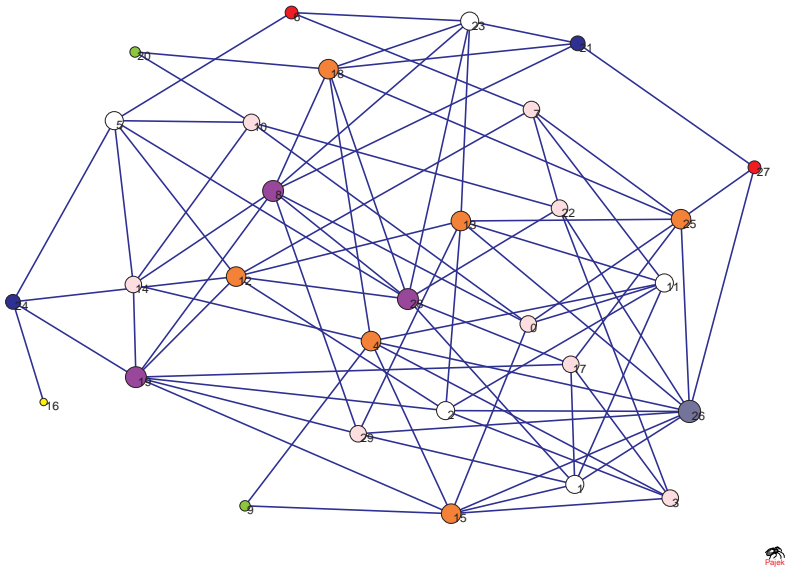}}
	\caption{A random network with $N=30$ and $<d>=5$. Nodes 16 and 24 form the ``trap" described in Section IV C.}
\end{figure*}

Though a variety of scenarios are possible, by far the most common occurrence is a two-node trap involving an outer node with degree one linked to another (inner) node with a higher degree, which in turn is linked to other high degree nodes. A network exhibiting such a structure is shown in Fig.\ 6. This is a random network characterized by $N=30$ and $<d>=5$. The trap is formed by nodes 24 and 16. Even from the beginning, when the queues are short, node 16 receives less traffic than any other node, thus having a shorter queue on average. This does not help its inner companion though, since it will be more likely to send its particles towards the outer node only to receive them back. When the queue length of node 24 is large, it will not receive any particles from the rest of the network until its queue length decreases. Thus we have periods of time when the transport between the two parts of the network is interrupted. As long as the queue lengths are small and their fluctuations large compared to their averages, these periods are finite. But at some point, due to a large statistical fluctuation of the queue length of node 24, particles that cannot reach their destination start accumulating on both sides. This can give rise to various behavior patterns depending on network topology and load. The simplest pattern occurs when the outer node consistently maintains a shorter queue than the nodes in the rest of the network, while the inner node maintains a longer queue. If the routing rule rigorously points towards the neighbor with the shortest queue, this effectively blocks the exchange of particles between the two nodes and the rest of the network. The inner node will only send particles toward its outer companion, which can only return them back. The nodes in the main part of the network to which the inner one is connected will always find other neighbors with shorter queues and send the particles their way. On both sides, particles that cannot find their destination keep accumulating. This eventually increases the destination finding time even for particles that can find their destination, due to the longer waiting periods in queues.

The scenario is illustrated in Figs.\ 7 and 8, which both pertain to the network in Fig.\ 6. Fig.\ 7 shows plots of the fractions of the total number of particles sitting in the queues of the various nodes as a function of time. This is a case when the jamming of the network occurs some $5\times 10^4$ time steps into the simulation. As can be seen, after the onset of jamming the queue length of node 16 is consistently shorter than the queue lengths of all other nodes, while the queue length of node 24 is consistently longer. Plots of the number of particles having a given destination versus time are shown in Fig.\ 8. Here we see how a large fluctuation of the queue length of node 24 prevents the particles on the main network from reaching the two nodes and triggers an initially unstable accumulation of particles seeking nodes 16 and 24. A few thousand time steps later, the network passes a tipping point and these particles start accumulating at a higher rate, driving up the number of particles with other destinations as well. More complicated patterns are possible, especially over large time scales, when either the queue length of the outer node becomes larger than that of the inner node or when, as jamming progresses, the queue length of some nodes in the rest of the network become larger.
\begin{figure*}
	\scalebox{0.5}{\includegraphics{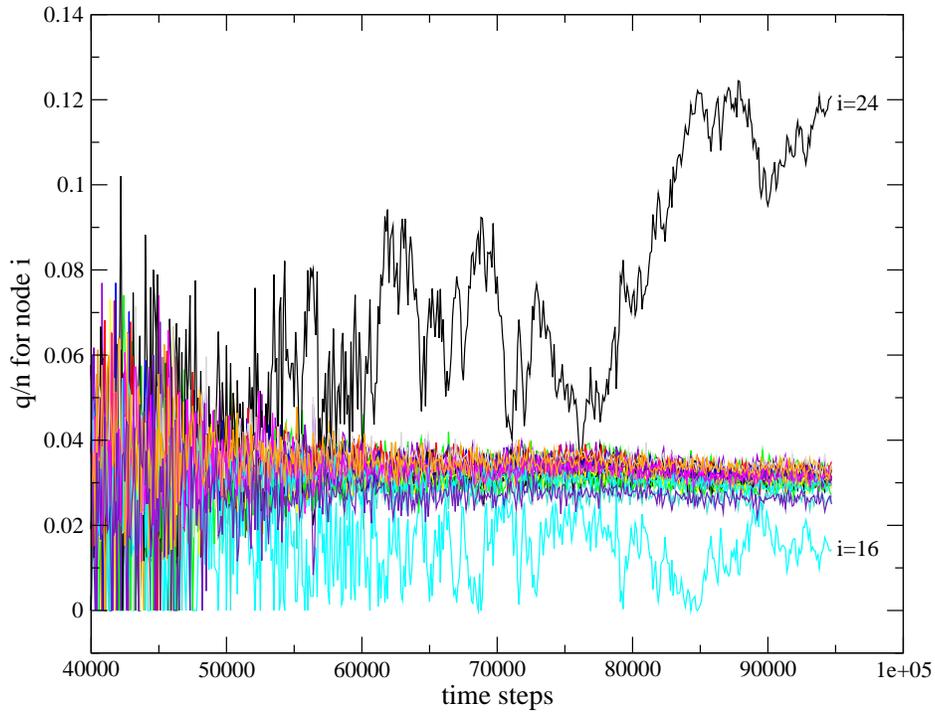}}
	\caption{(Color online) Plot of the number of particles $q_i$ sitting in the queue of node $i$ divided by the total number of particles on the network $n$ versus time. Each curve has a different color and corresponds to a different queue.}
\end{figure*}

Other types of structures are also prone to trap formation, but they are less likely to occur. These include traps consisting of a chain-like structure connected to a higher degree node, two or more nodes of degree one (outer nodes) connected to the same high degree node, as well as traps involving nodes of degrees higher than one (for example a triangle connected only at one vertex with the rest of the network or a chain between two high degree nodes). In all these cases particle exchange may take place sporadically between these subsets and the rest of the network.

Now the main reason for the existence of an optimum value of $\beta$ becomes apparent. By retaining some stochasticity in the routing rule, the onset of the transport traps can be avoided since particles are sufficiently likely to be sent even towards neighbors that have longer queues than others.
\begin{figure*}
	\scalebox{0.5}{\includegraphics{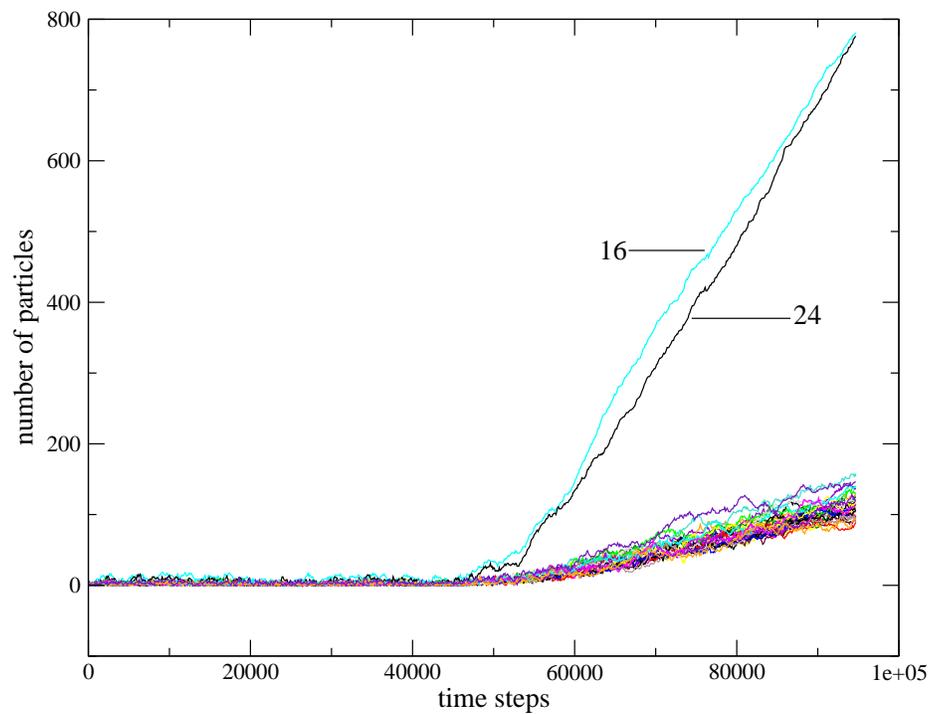}}
	\caption{(Color online) Plot of the number of particles on the network that have a given node as destination versus time. Each curve has a different color and corresponds to a different destination node.}
\end{figure*}

Depending on the topology of the network around the critical nodes, traps can occur virtually from the beginning, without the network ever reaching a steady state, or they can be triggered by a large statistical fluctuation at a later time. Indeed, the possibility of jamming after an initial period of steady transport is a characteristic of networks operating under (almost) deterministic congestion aware routing. In all other cases, networks either reach a steady state that lasts indefinitely, or start jamming from the beginning. The connection between jamming start time and the details of network structure seems to be extremely complicated. Furthermore, since simulations can only be run for finite periods of time, we cannot identify all networks that jam due to the formation of transport traps. However, from the point of view of rigid congestion aware routing it seems safe to distinguish between two types of networks: those that are ``structurally fit" for it and able to bear as high loads as at intermediate degrees of congestion awareness, and those that are prone to the formation of traps (due to the presence of structural features like those discussed earlier in this section) and end up jamming under lower loads than in the case of a less rigid routing. At the same number of nodes and average degree, scale-free networks are more likely to exhibit transport traps than random networks.

\section{Conclusions}

\noindent We use a simple model to study the behavior of network transport in the case of routing based on local information with various degrees of congestion awareness, ranging from random diffusion to the extreme case of rigid congestion-gradient driven flow. The degree of congestion awareness is controlled by a single scalar parameter. The average transport capacity for networks with a given set of topology and transport parameters is characterized by the critical load under which half of these networks are jamming. At the average connectivity we have used, random networks are more robust against jamming than scale-free networks with the same number of nodes and average degree. However, in light of the results presented in \cite{ParkLaiZhaoYe}, it is possible that scale-free networks become more robust once a certain critical value of the average connectivity is exceeded. Regardless of topology, the critical load decreases as the number of nodes increases, which means that the critical load per node approaches zero in the limit of large number of nodes. Consequently, jamming seems to be unavoidable in sufficiently large networks with transport based on local information only.

A somewhat counterintuitive result is the existence of an optimum value of the congestion awareness parameter. Below this value, transport capacity increases with the degree of congestion awareness, and reaches its maximum at the optimum value. The increase is due to the fact that particles are more likely to avoid sitting in the queues of the busiest nodes (``hubs") for extended periods of time. A high degree of congestion awareness does, however, have some unwanted effects which eventually lead to a decrease in transport capacity. We show that this decrease is mainly due to the occurrence of transport traps, which prevent particle exchange between parts of the network. These traps are formed by sets of low degree nodes (``stubs") connected to nodes with higher degrees. The optimum value of the congestion awareness parameter arises from the interplay between the effects that tend to increase the transport capacity and those that tend to decrease it. The overall lower robustness of the scale-free networks is due to the fact that a power-law distribution of the node degrees allows more high degree nodes (which are responsible for jamming in the case of low congestion awareness) and also more low degree nodes (responsible for jamming in the case of high congestion awareness) compared to the binomial distribution that characterizes the random networks.

We have also shown that a betweenness centrality measure similar to the one defined in \cite{Guimera} provides an essentially exact description of the statistics of network transport in the case of random diffusion. In addition, we describe an algorithm for the fast computation of this betweenness measure. Furthermore, the betweenness proves to be a useful tool in the analysis of network transport under congestion aware routing as well.

\begin{acknowledgments}

The authors gratefully acknowledge support from A.\ Williams of the Air Force Research Laboratory, Information Directorate, under contract No.\ FA8750-04-C-0258. BD, YY, and KEB also acknowledge support from the NSF through grant No.\ DMR-0427538. ZT was supported by DOE contract No.\ W-7405-ENG-36.

\end{acknowledgments}


\begin{thebibliography}{99}

\bibitem[1]{NewmanSIAM}
M.\ E.\ J.\ Newman, SIAM Review {\bf 45}, 167 (2003).

\bibitem[2]{WattsStro}
D.\ J.\ Watts and S.\ H.\ Strogatz, Nature {\bf 393}, 440 (1998).

\bibitem[3]{HolmeCong}
P.\ Holme, Advances in Complex Systems {\bf 6}, 163 (2003).

\bibitem[4]{EcheniquePRE}
P.\ Echenique, J.\ G\'omez-Garde\~nes, and Y.\ Moreno, Phys.\ Rev.\ E {\bf 70}, 056105 (2004).

\bibitem[5]{EcheniqueEL}
P.\ Echenique, J.\ G\'omez-Garde\~nes, and Y.\ Moreno, Europhys.\ Lett.\ {\bf 71}, 325 (2005).

\bibitem[6]{Anghel-ea}
M.\ Anghel, Z.\ Toroczkai, K.\ E.\ Bassler, and G.\ Korniss, Phys.\ Rev.\ Lett.\ {\bf 92}, 058701 (2004).

\bibitem[7]{Lopez-ea}
E.\ Lopez, S.\ V.\ Buldyrev, S.\ Havlin, and H.\ E.\ Stanley, Phys.\ Rev.\ Lett.\ {\bf 94}, 248701 (2005).

\bibitem[8]{Korniss-ea}
G.\ Korniss, M.\ B.\ Hastings, K.\ E.\ Bassler, et al., Physics Letters A {\bf 350}, 324-330 (2006).

\bibitem[9]{Guimera}
R.\ Guimer\`a, A.\ D\'{\i}az-Guilera, F.\ Vega-Redondo, A.\ Cabrales, and A.\ Arenas,
Phys.\ Rev.\ Lett.\ {\bf 89}, 248701 (2002).

\bibitem[10]{BarabRMP}
R.\ Albert and A.-L.\ Barab\'asi, Reviews of Modern Physics {\bf 74} (2002).

\bibitem[11]{BarabLinked}
A.-L.\ Barab\'asi, \emph{Linked}, (Perseus Publishing, 2002).

\bibitem[12]{NewmanPRE}
M.\ E.\ J.\ Newman, Phys.\ Rev.\ E {\bf 64}, 016132 (2001).

\bibitem[13]{BarabScience}
A.-L.\ Barab\'asi and R.\ Albert, Science {\bf 286}, 509 (1999).

\bibitem[14]{ZJHaas}
Z.\ J.\ Haas et al.\ editors, \emph{Special Issue on Wireless Ad Hoc Networks}, IEEE J.\ on Selected Areas in Communications {\bf 17}, No.\ 8 (1999).

\bibitem[15]{ZhaoLaiParkYe}
L.\ Zhao, Y.-C.\ Lai, K.\ Park, and N.\ Ye, Phys.\ Rev.\ E {\bf 71}, 026125 (2005).

\bibitem[16]{TB}
Z.\ Toroczkai and K.\ E.\ Bassler, Nature {\bf 428}, 716 (2004).

\bibitem[17]{TKBHK}
Z.\ Toroczkai, B.\ Kozma, K.\ E.\ Bassler, N.\ W.\ Hengartner, and G.\ Korniss, arXiv:cond-mat/0408262, v1, 12 Aug 2004.

\bibitem[18]{Erdos}
P.\ Erd\H{o}s and A.\ R\'enyi, Publ.\ Math.\ (Debrecen) {\bf 6}, 290 (1959);
P.\ Erd\H{o}s and A.\ R\'enyi, Bull.\ Inst.\ Int.\ Stat.\ {\bf 38}, 343 (1961).

\bibitem[19]{ParkLaiZhaoYe}
K.\ Park, Y.-C.\ Lai, L.\ Zhao, and N.\ Ye, Phys.\ Rev.\ E {\bf 71}, 065105(R) (2005).

\bibitem[20]{Newman-e}
M.\ E.\ J.\ Newman, arXiv:cond-mat/0309045, v1, 1 Sep 2003.

\bibitem[21]{Estrada}
E.\ Estrada and J.\ A.\ Rodr\'{\i}guez-Vel\'azquez, Phys.\ Rev.\ E {\bf 71}, 056103 (2005).

\bibitem[22]{Latora-e}
V.\ Latora and M.\ Marchiori, arXiv:cond-mat/0402050, v1, 2 Feb 2004.

\bibitem[23]{Brajendra}
B.\ K.\ Singh and N.\ Gupte, Phys.\ Rev.\ E {\bf 71}, 055103(R) (2005).

\bibitem[24]{SMWformula}
G.\ H.\ Golub and C.\ F.\ Van Loan, \emph{Matrix computations}, 3rd ed. (Johns Hopkins University Press, 1996).

\bibitem[25]{Allen}
O.\ Allen, \emph{Probability, Statistics and Queueing Theory with Computer Science Application}, 2nd ed. (Academic Press, New York, 1990).

\end{thebibliography}

\end{document}